\documentstyle[preprint,aps]{revtex}
\begin{document}
\draft
\title{
The Fermi surface of  underdoped high-$T_c$\\
 superconducting cuprates
}
\author{Xi Dai,$^a$  Zhao-bin Su,$^a$ and Lu Yu$^{b,a}$}
\address{
$^{a}$ Institute of Theoretical Physics, Academia Sinica,
P.O. Box 2735, Beijing 100080, China\\
$^b$ International Center for Theoretical Physics,
P.O. Box 586, Trieste 34100, Italy
}
\maketitle

\begin{abstract}
The coexistence of $\pi$-flux state and d-wave RVB state is considered in
this paper within the slave boson approach. A critical value of doping
concentration $\delta_c$ is found,  below which
the coexisting  $\pi$-flux   and d-wave RVB state is favored in
energy. The pseudo Fermi surface of spinons and the physical electron spectral
function are calculated. A clear Fermi-level
 crossing is  found along the (0,0) to ($\pi$, $\pi$) direction,
but no such crossing is detected along the ($\pi$, 0) to ($\pi$, $\pi$)
direction. Also, an   energy gap of d-wave symmetry  appears at the Fermi
level in
our calculation. The above results are in agreement with the
angle-resolved photoemission experiments which indicate at  a d-wave pseudo-gap
and a half-pocket-like Fermi surface in underdoped cuprates.
\end{abstract}

\vspace*{.8cm}

{\bf Category}: Ca2

{\bf Contact Author}: Xi Dai

{\bf Mailing Address}: Institute of Theoretical Physics,
 P.O. Box 2735, Beijing 100080,\\
\hspace*{4cm}  P. R. of China

{\bf E-mail}: daix@sun.itp.ac.cn

\newpage

\centerline{\bf I. INTRODUCTION}
\vspace{.5cm}

A great deal  of attention has been focused on the studies of
the electronic structure
of the high $T_c$ superconducting cuprates (HTSC) in the past
 decade\cite{dag,kam,bre}.
In the undoped case, the charge mobility is suppressed by the strong on-site
Coulomb repulsion and the spin-${1 \over 2}$ antiferromagnetic Heisenberg
model is
believed appropriate  to describe the electronic properties of these
materials\cite{kam}.
In the opposite limit, when the cuprates are overdoped in the copper-oxide
plane, the HTSC is metallic and the observed electronic structure is
in good agreement with
 predictions of the local density approximation (LDA) calculations\cite{pic}.
 In-between
these two limits, many interesting phenomena appear including
superconductivity. The evolution of the electronic structure from one limit
to  another should  play a key role for  the understanding of
numerous  anomalous properties of
HTSC discovered in this doping range.

When slightly doped with holes in the antiferromagnetic Cu-O plane, the Fermi
surface of the mobile holes is expected to be  small pockets centered at
$(\pm\pi /2,\pm\pi /2)$. On the contrary, in the overdoped case the Fermi
surface
should  be a large one centered at $(\pi,\pi)$ as reported  by both
 angle-resolved photo-emission spectroscopy (ARPES) experiments\cite{shen}
and the
LDA calculations\cite{pic}. However,  the crossover from one limit  to another
 is not well understood yet.

~~~~The recent ARPES experiments\cite{mar,loe,ding} on  underdoped
samples seem to indicate that the Fermi surface of these compounds is
probably half-pocket-like, {\it i.e.} a small pocket near $\Sigma~(\pi/2,
\pi/2)$,
but lacking its outer part
in the reduced Brillioun zone scheme ( see the dashed line in Fig. 3).  The data
of  Marshall {\it et al.}\cite{mar,loe} show a well defined
Fermi-level crossing along the (0, 0) to ($\pi$, $\pi$) direction, but
no such crossing has been detected along the ($\pi$, 0) to ($\pi$, $\pi$)
direction. This result  differs substantially
from that of the optimally
doped samples which exhibit a large Fermi surface with well defined
Fermi-level crossing  in both (0, 0) to ($\pi$, $\pi$) and
($\pi$, 0) to ($\pi$, $\pi$) directions.

~~~~Another puzzling phenomenon in underdoped cuprates is the
presence of a pseudo-gap  in the normal state. Both Stanford\cite{loe} and
Argonne\cite{ding} groups have shown  strong evidence for the
suppression of the density
of states near the Fermi level  indicating at the formation
of the pseudo-gap. Several different measurements,
including specific
heat and thermal power\cite{lor,tal}, Hall coefficient\cite{hwa}, and
transport\cite{bat}, as well as nuclear magnetic
resonance (NMR)\cite{tak} and  neutron scattering experiment
\cite{ros}, have also shown
the presence of the pseudo-gap in underdoped cuprates.
 The symmetry of the pseudo-gap has been found
d-wave like\cite{loe,ding} with a  maximum value in the ($\pi$, 0) to
($\pi$, $\pi$) direction,
 and it is very much reduced  in the (0, 0) to ($\pi$, $\pi$) direction.
This result is also consistent with earlier data of Marshall
 {\it et al.}\cite{mar}, where dramatical changes
have been observed for the spectral density near (0, $\pi$).

Theoretically, the origin of the pseudo-gap can be ascribed
either to short range AF fluctuations
in the weak coupling approach\cite{ks} or to
d-wave pairing of spinons\cite{fuk,wen} in the  resonant valence bond (RVB)
 state\cite{and}.
Due to the SU(2) symmetry at half filling, the d-wave and s+id wave\cite{kot}
RVB states, as well as $\pi$ flux state\cite{mars}
 are degenerate\cite{aff}. Away from half
filling, the commensurate flux phase (CFP)\cite{led}, the  staggered
 flux phase (SFP)\cite{har}
and the d-wave RVB state\cite{zhang} are competing candidates
for the ground
state. Further study shows that the CFS is  favoured only in the limit
$t \ll J$\cite{led}, which is not the actual case in cuprates. Zhang\cite{zhang}
 has studied the
competition and possible coexistence of the SFP and d-wave RVB state, and
he concluded that SFP is unstable against d-wave RVB state and there is
 no coexistence of these  two states  at any  non-half-fillings.
However,  Zhang's
conclusion was based on the renormalized mean-field theory which satisfies
the no-double occupancy constraint only on average. To improve
the situation, in one of our earlier papers\cite{ssy}, D.N. Sheng and two of
us (Z.B. Su and L. Yu) have reconsidered this issue using
a better treatment of the slave-boson mean field theory, proposed by
Nori, Abrahams and Zimanyi (NAZ)\cite{nor},
 transmuting the
holons into spinless Fermions via a 2D Jordan-Wigner transformation.
Using NAZ statistics transmutation trick it was found that the $\pi$-flux
 state can coexist with the s+id-wave RVB state
 below a critical doping $\delta_c$, while the d-wave state takes over beyond
that critical concentration.

Recently, a SU(2) slave boson theory  was proposed by Wen and Lee\cite{wen},
significantly improving the earlier $U(1)$ version, especially for the
underdoped systems. In their approach
the SU(2) symmetry is preserved away from  half-filling by introducing
two kinds of slave bosons. Considering the effective interaction between holons
and spinons they obtain  a small  Fermi Surface
 in underdoped regime.  We will elaborate further on their
results later, comparing them with ours.

In the present paper we develop a slave boson approach for the $t-J$ model,
with a more careful treatment of  the hard-core constraint for  the holons
 by transforming them  into fermions.
 Following NAZ\cite{nor}, a  statistical gauge field $A_{ij}$ is
introduced to implement this transmutation.
 The effective gauge field
seen by the holons will be a superposition of the $\pi$-flux contributed
by the spinons and the statistical gauge field reflecting the hard-core
nature of the holons. Using the equivalence of the $\pi$-staggered
and $\pi$-uniform flux, we find that the statistical gauge field can
substantially
compensate the $\pi$-flux. Then the effective flux acting on holons
satisfies
the commensurability condition which minimizes the holon free energy.
 The mean field phase diagram is obtained   and the coexisting
 d-wave RVB and  $\pi$-flux  state is found energetically more favourable
with respect to the pure d-wave sate in the underdoped region.
This confirms the earlier result\cite{ssy}, and disagrees with
 Zhang's conclusion\cite{zhang}, excluding the coexistence of these
two states. It appears to us that  the
main reason of this discrepancy is due to the fact that
 in Zhang's approach the holons and spinons
are not treated on an equal footing. The dynamics of holons is
basically ignored in his approach, leading only to the renormalization
of the hopping integral. On the contrary, in our approach, the full dynamics
of holons is  considered and the holons and spinons are
treated equally. ( We will elaborate more on the differences of these
two approaches in the concluding remarks.)
Furthermore,  we have calculated the
physical electron spectral function in the low doping range and have found the
half-pocket-like Fermi surface as well as  the pseudo-gap in the same doping
range. We emphasize that the pseudo-gap obtained in our approach  opens
right at the Fermi surface and therefore it is consistent with the ARPES
experiments.
Thus our calculation can very well explain the APRES results  without
invoking the additional  (not physical) SU(2) symmetry away from half-filling.

The rest of the paper is organized as follows:
The mean field theory is given in Sec. II, whereas the spectral function for
the physical electron is presented in Sec. III.
Finally, we make a few concluding remarks in Sec. IV.

\vspace{4ex}
\vspace{10mm}

\centerline{\bf II. MEAN FIELD THEORY OF THE  $t-J$ MODEL}

\vspace{.5cm}

   We begin with the slave boson representation of the $t-J$ model
in which the physical electron destruction  operator $c_{i\sigma}$ is
represented
in terms of the pseudo-fermion operator $f_{i\sigma}$ and the slave boson
operator
$b_i^+$ as
$c_{i\sigma}=f_{i\sigma}b_i^+$. The partition function reads as :

\begin{equation}
Z=\int
\prod_{i\sigma}{D[f_{i\sigma}^+(\tau)]D[f_{i\sigma}(\tau)]D[b_{i}^+(\tau)]
D[b_{i}(\tau)]}~\exp[\displaystyle-\int_0^{\beta}Ld\tau]
\end{equation}

with
$$
L=\sum_{i\sigma}f_{i\sigma}^+(\tau)(\frac{\partial}{\partial \tau}-\mu_2)
f_{i\sigma}(\tau)+\sum_ib_i^+(\tau)(\frac{\partial}{\partial
\tau}-\mu_1)b_i(\tau)
+N\mu_1\delta+N\mu_2(1-\delta)
~~~~~~~~~~~~~~~~~~~~~~~~~~~~~~~~
$$
$$
-{1 \over 2}\sum_{\langle ij
\rangle}(\tau_{ij}+t\;b_i^+b_j)(\tau_{ji}+t\;b_j^+b_i)
-{t^2 \over 2} \sum_{\langle ij \rangle}b_i^+b_ib_j^+b_j+{t^2 \over
2}\sum_ib_i^+b_i
$$
\begin{equation}
-{1 \over 2}\sum_{\langle ij \rangle}\Delta_{ij}^+\Delta_{ij}+
t'\sum_{(ii')}f_{i\sigma}^+f_{i'\sigma}b_ib_{i'}^+,
\end{equation}
where we define
$$
\tau_{ij}=\sum_{\sigma}f_{i\sigma}^+f_{j\sigma}, ~~~~~~~~~~~~~~~~~
\Delta_{ij}=\sum_{\sigma}\sigma f_{i\sigma}f_{j\bar{\sigma}}.
~~~~~~~~~~~~~~~~~~~~~~~~~~~~~~~~~~~~~~~~~~
$$
The chemical potentials for the boson $\mu_1$ and the fermion $\mu_2$
are chosen to satisfy the global constraints  $n_f+n_b$ =1 and $n_b=\delta$,
with $\delta$ as the doping concentration. We take the exchange integral
$J = 1$, while $t$, $t'$ are the nearest neighbor, and next nearest hopping
terms,
respectively. Hereafter $\langle ij \rangle$ means nearest neighbor, while
(ii') means
next nearest summation. $N$ is the total number of sites, and $\beta$ is
the inverse temperature.

We introduce the Hubbard-Stratonovich (HS) transformation:

$$
Z=\int \prod_{i\sigma}D[f_{i\sigma}^+]D[f_{i\sigma}]
D[b_{i}^+]D[b_{i}]
\prod_{\langle ij \rangle}D[\chi_{ij}^*]D[\chi_{ij}]D[D_{ij}^*]D[D_{ij}]
$$
\begin{equation}
\cdot \prod_{(ii')}
D[\rho_{ii'}^*]D[\rho_{ii'}]~\exp[\displaystyle-\int_0^{\beta}L'(\tau)d\tau]
\end{equation}

with
$$
L'(\tau)=\sum_{i\sigma}f_{i\sigma}^+(\tau)(\frac{\partial}{\partial \tau}-\mu_1)
f_{i\sigma}(\tau)+\sum_ib_i^+(\tau)(\frac{\partial}{\partial \tau}-\mu_2)
b_i(\tau)+{1 \over 2}\sum_{\langle ij \rangle}\chi_{ij}^*(\tau)\chi_{ij}(\tau)
~~~~~~~~~~~~~~~~~~~~~~~~~
$$
$$
+{1 \over 2}\sum_{\langle ij
\rangle}D_{ij}^*(\tau)D_{ij}(\tau)-t'\sum_{(ii')}\rho_{ii'}^*(\tau)
\rho_{ii'}(\tau)-{1 \over 2}\sum_{\langle ij
\rangle\sigma}\left[\chi_{ij}^*(\tau)
f_{i\sigma}^+(\tau)f_{j\sigma}(\tau)+h.c.\right]
~~~~~~~~~~~~~~~~~~~~~~~~~~
$$
$$
-{1 \over 2}\sum_{\langle ij \rangle
\sigma}\left[D_{ij}^*(\tau)\Delta_{ij}+h.c.\right]
-{t \over 2} \sum_{\langle ij \rangle}(\chi_{ij}^*b_i^+b_j+h.c.)
+{t'}\sum_{(ii')\sigma}\left[\rho_{ii'}^*f_{i\sigma}^+f_{i'\sigma}+
h.c.\right]
~~~~~~~~~~~~~~~~~~~~~~~
$$
\begin{equation}
+{t' }\sum_{(ii')}\left[
\rho_{ii'}b_{i'}^+b_i+h.c.\right]-{t^2 \over 2}\sum_{\langle ij
\rangle}b_i^+b_ib_j^+b_j+
{t^2 \over 2}\sum_ib_i^+b_i
+\mu_1\sum_i\delta+\mu_2\sum_i(1-\delta).
\end{equation}

In the saddle point approximation of the above action, we treat the
amplitudes of
the HS fields $D_{ij}$,
$\chi_{ij}$ and $\rho_{ii'}$ as site- and time-independent with some phase
factors, namely,
$$
D_{ij}(\tau)=\Delta\;\eta_{ij}, ~~~~~ \chi_{ij}(\tau)=p\;\exp(i\phi_{ij}),
 ~~~~~~\rho_{ii'}(\tau)=\rho
$$
with
$$
\eta_{ij} = \left\{
\begin{array}{l}
1~(\hat{j} = \hat{i}\pm \hat{x}),\\
-1~ (\hat{j} =
\hat{i}\pm \hat{y}),
\end{array}
\right.
$$
while $\phi_{ij}$ is the  phase factor in the Affleck-Marston flux
phase\cite{mars}
to be
described in detail later.

We then obtain the mean-field Hamiltonian per site  as:

$$
H_{MF}=H_c+H_s+H_h,
$$
$$
H_c={1 \over 2}\Delta^2+{1 \over 2}p^2+\mu_2(1-\delta)+\mu_1\delta-
{t^2 \over 2}\delta (\delta -1)
+t'|\rho|^2,
$$
$$
H_s=-\frac{p}{2N}\sum_{\langle ij
\rangle\sigma}(e^{i\phi_{ij}}f_{i\sigma}^+f_{j\sigma}
+h.c.)
-\frac{\Delta}{2N}\sum_{\langle ij \rangle\sigma}\left[\eta_{ij}\sigma
f_{i\sigma}^+
f_{j\bar{\sigma}}^+ + h.c. \right]
$$
$$
+\frac{t'\rho}{ N}\sum_{(ii')\sigma}(f_{i\sigma}^+f_{i'\sigma}
+h.c.)-\frac{\mu_2}{N}\sum_{i\sigma}f^+_{i\sigma}f_{i\sigma}
$$
\begin{equation}
H_h=-\frac{t p}{2N}\sum_{\langle ij \rangle}(e^{i\phi_{ij}}b_i^+b_j+h.c.)+
\frac{t'\rho}{N}\sum_{(ii')}(b_i^+b_{i'}+h.c.)
-\frac{\mu_1}{N}\sum_ib_i^+b_i.
\label{hol}
\end{equation}

We introduce  a statistical gauge field operator
$A_{ij}$ to account for the hard-core nature of the
holons, namely,
$b_i^+b_j=exp(iA_{ij})h_i^+h_j$, with $h_i^+$, $h_j$ as fermion operators.
In the mean field approximation, the statistical field $A_{ij}$ is considered
as a c-number, but it must obey
the  constraint:
$$\sum_{plaquette}A_{ij}=2\pi(2q+1)\delta$$
 with $q$ as an integer.,
This constraint  can be satisfied by introducing an additional
 uniform flux attached to the holons. Then
the effective mean field Hamiltonian for holons reads as :

\begin{equation}
H_h=-\frac{tp}{2N}\sum_{\langle ij
\rangle}\left[e^{i(\phi_{ij}+A_{ij})}h_i^+h_j+h.c.\right]
-\frac{t'\rho}{2N}\sum_{(ii')}\left[e^{i(\phi_{ii'}+A_{ii'})}h_i^+h_{i'}+h.c
.\right]
-\frac{\mu_1}{N}\sum_ih_i^+h_i.
\end{equation}

The effective flux that acts on the holons is then :
$$
\Phi_h=\sum_{plaquette}(A_{ij}+\phi_{ij})+2\pi k=2\pi\delta+4\pi q\delta+2\pi k
+\pi,
$$
where $k$ is an integer.
If $\delta=(2p_1-1)/4k_1$, we can choose proper $q_1$ and $k_1$ to satisfy the
commensurability condition $\Phi_h=2\pi\delta$ which minimizes the total
free energy of holons .
In the present calculation we choose $k_1=25$ and $p_1=1,2......,25$, so the
above commensurability condition is always satisfied.
Thus we transform the holon Hamiltonian into a simple form
equivalent to the problem of noninteracting lattice fermions moving in a
 uniform magnetic field. We then diagonalize this mean field Hamiltonian of
holons
by solving the well-known Harper's equation.

For the spinon Hamiltonian we choose a proper gauge  shown in Fig. 1,
where $\phi_{ij}={\pi \over 4}$ if $R_{ij}$ is along the  arrow shown in the
figure and  $\phi_{ij}=-{\pi \over 4}$  in the opposite case.
 Then we divide the original lattice
into two sublattices,labeled by 1 and 2. The spinon Hamiltonian written in
the $k$-space is found to be:

$$
H_s=- {1 \over N}{\sum_{k\sigma}}'(e(k)+\mu_2)f_{1k\sigma}^+f_{1k\sigma}
   -{1 \over N}{\sum_{k\sigma}}'(e(k)+\mu_2)f_{2k\sigma}^+f_{2k\sigma}
$$
\begin{equation}
-{1\over N}{\sum_{k}}'
\frac{p}{\sqrt2}\left[(\gamma_++i\gamma_-)f_{1k\sigma}^+f_{2k\sigma}
+h.c.\right]
+{\sum_k}' \Delta \gamma_- \left[f_{1k\uparrow}^+f_{2-k\downarrow}^+
+h.c.\right],
\end{equation}
where $\sum'$ means  summation over the reduced
Brillouin zone with  reciprocal vector $Q=(\pi,\pi)$,
$$\gamma_{\pm}=(\cos k_x\pm\cos k_y),~~~~~~~~~~~~~
e(k)=2t'\rho \left[\cos (k_x-k_y)+\cos (k_x+k_y)\right].
$$

The above Hamiltonian can be diagonalised by the Bogoliubov transformation,
and we obtain the quasi-particle energy as $\pm E_{ks}$, and
with $s=\pm 1$:

\begin{equation}
E_{ks}=\left\{\left[ sp~\sqrt{\displaystyle\frac{\cos^2(k_x)+\cos^2(k_y)}
{2}}-\mu_2-e(k)\right]^2+\Delta^2
\gamma_-^2\right\}^{\displaystyle 1/2}.
\end{equation}

Now we  calculate the spinon and holon free energy for  given values
of $p$, $\delta$, $\rho$, $\mu_1$ and $\mu_2$. The parameters are chosen
as $t = 2$, $t'=0.2$, and the free energy is calculated at temperature
$T=0.05$. We minimize the total free energy by varying the order parameters $p$,
$\Delta$, $\rho$ and $\mu$ to obtain the mean field phase diagram
as a function of doping concentration. The spinon dispersion obtained
by our approach shows that the pseudo-gap  opens up right at the Fermi
level which vanishes along the line $k_x=\pm k_y$ reflecting the d-wave
symmetry.

As shown in Fig. 2, the coexistence of $\pi$-flux and
d-wave RVB state is favored in energy in the low doping range, while
in the  high doping range the pure d-wave RVB state takes over.
The critical value of the doping concentration is found to be near
0.3. We believe  this phase diagram is  correct only qualitatively, since
the fluctuations
around the saddle point have been ignored in our present study. This
calculation has confirmed the earlier result\cite{ssy}. The new
ingredient here is to introduce the next nearest neighbor hopping $t'$ which
turns out to be essential for obtaining the correct shape of the spectral
function.
 Like in the CFP\cite{led}, the mean field
treatment of the commensurability condition $\Phi_h= 2\pi\delta$
for holons breaks the parity and time-reversal symmetry at arbitrary filling.
However, this is an artifact due to the mean field treatment.

\vspace{4ex}
\vspace{10mm}

\centerline{\bf  III. ELECTRON SPECTRAL FUNCTION}
\vspace{.5cm}

Now we calculate the spectral function for physical electrons.
It is more convenient to use the boson representation for holons
instead of the fermion representation
because we can not fully restore the Bose statistical property of the holon
operator
in the mean field approximation when the operator $A_{ij}$ is treated as a
c-number. The boson representation for holons is a further approximation
compared with the mean field treatment in the preceding Section.
However, we do use the coexisting $\pi$-flux and d-wave RVB order
parameters obtained there to calculate the electron spectral function.
Presumably, the main physical consequences of the coexisting phase
are still kept.

First diagonalize the
holon Hamiltonian in this boson representation. After Fourier transformation
Eq.( \ref{hol}) reads as :

\begin{equation}
H_h=- {1 \over N}{\sum_{k}}'(e(k)+\mu_1)b_{1k}^+b_{1k}
   -{1\over N}{\sum_{k}}'(e(k)+\mu_1)b_{2k}^+b_{2k}
-{1\over N}{\sum_{k}}'\frac{tp}{\sqrt2}\left[(\gamma_++i\gamma_-)b_{1k}^+b_{2k}
+h.c.\right],
\end{equation}
where the indices 1,2 represent the two sublattices defined above.

This Hamiltonian can be diagonalised as:

\begin{equation}
H_h=\frac{1}{N}{\sum_q}'E_b(q)(\beta_q^+\beta_q-\alpha_{q}^+\alpha_{q})-
\frac{\mu_1+e(q)}{N}{\sum_q}'(\beta_q^+\beta_q+\alpha_{q}^+\alpha_{q})
\end{equation}
with
$$
E_b(q)=\frac{pt}{\sqrt2}\sqrt{\cos^2(q_x)+\cos^2(q_y)},
$$
which describes two kinds of free bosons labeled by $\alpha$ and
$\beta$. At low temperatures, bosons  occupy the low energy states.
So we need to  consider only  the bottom of the lower band (the $\alpha$ band)
near $\Gamma$ point (0, 0). Expanding
$E_b(q)$ near (0,0), we  obtain the effective holon Hamiltonian as:

\begin{equation}
H_h=\sum_{q}(\frac{q^2}{2m_B}-\mu_B)\alpha_q^+\alpha_q,
\end{equation}
where $\mu_B$ is the renormalized boson chemical potential, and
$$
\frac{1}{2m_B}=\frac{ tp}{8}+2t'\rho
$$
The physical electron Green's function is:
\begin{equation}
G(k,\tau)=-\langle T_{\tau}(c_{k\sigma}(\tau)c_{k\sigma}^+(0))\rangle,
\end{equation}
where
$$
c_{k\sigma}={1 \over N}\sum_{i}c_{i\sigma}\exp(-ik\cdot R_i)
={\sqrt2 \over 2} (c_{1k\sigma}+c_{2k\sigma}).
$$
Hence
$$
G(k,\tau)=- {1\over 2}\sum_{\alpha,\beta}\langle T_{\tau}(c_{\alpha
k\sigma}(\tau)c_{\beta
k\sigma}^+(0))\rangle
$$
\begin{equation}
=-{1\over 2}\sum_{\alpha,\beta,q}\langle T_{\tau}(f_{\alpha
k+q\sigma}(\tau)f_{\beta
k+q\sigma}^+(0))\rangle\langle T_{\tau}(b_{\alpha q}^+(\tau)b_{\beta
q}(0)\rangle.
\end{equation}
We  see that the physical electron Green's function in the slave-boson
approximation is nothing but the fermion-boson polarization function.
The spectral weight is thus:
\begin{equation}
{\rm Im}~G(k,\omega)=-\frac{1}{2N}{\sum_{\alpha \beta q \nu}}'S_{\alpha\nu}(k+q)
S^*_{\beta\nu}(k+q)D_{\alpha 1}^*(q)D_{\beta 1}(q)\delta(\omega+\omega_q
-\epsilon_{k+q}^\nu)\left[ n_B(\omega_q)+n_f(\epsilon_{k+q}^\nu)\right],
\label{spect}
\end{equation}
where  $S$ and $D$ are the transformation matrices used by us
to diagonalize $H_s$ and $H_h$, respectively.  The first term of Eq.
(\ref{spect})
contributes a quasiparticle like peak centered at $\epsilon_k^\nu$$-\mu_B$,
being called the coherent
part. The second term represents a featureless incoherent part extended over
a wide range. The density of occupied states  probed by  ARPES is
Im$G(k,\omega)
n_f(\omega)$. If we  consider only  the holon states with low energies as
discussed
above, we  find that the total  spectral weight of the
coherent part, detected by ARPES equals:

$$
\frac{1}{2N}{\sum_{\alpha\beta q\nu}}'\int d\omega S_{\alpha\nu}(k+q)
S^*_{\beta\nu}(k+q)D_{\alpha 1}^*(q)D_{\beta 1}(q)\delta(\omega+\omega_q
-\epsilon_{k+q}^\nu)n_B(\omega_q)n_f(\omega)
$$
$$
\approx\frac{1}{2N}{\sum_{\alpha\beta q\nu}}'S_{\alpha\nu}(k)
S^*_{\beta\nu}(k)D_{\alpha 1}^*(0)D_{\beta 1}(0)n_B(\omega_q)
n_f(\epsilon_{k})
$$
\begin{equation}
=\frac{1}{4N}{\sum_{\alpha\beta\nu}}'S_{\alpha\nu}(k)
S^*_{\beta\nu}(k)\delta~n_f(\epsilon_k)
={1 \over 2}<f_{k\sigma}^+f_{k\sigma}>\delta=\frac{1}{2}n_s(k)\delta
\end{equation}
Since the spectral weight itself is proportional to
 $n_s(k)$,  we  conclude that the shape of the  Fermi surface is
approximatively determined by the spinon momentum distribution  $n_s(k)$.
 The width of the peak is roughly proportional to  $T^{1/2}$,
which  is very similar to Lee and Nagaosa's\cite{ln} result for
the uniform RVB phase.

Fig. 4 presents  the momentum distribution of spinons
in different directions (as defined in Fig. 3) within the slave boson approach.
We  see that the curve $\alpha$ which is along the direction (1,1), {\it i.e.}
from (0, 0) to $(\pi, \pi)$,
 shows a clear step in the momentum distribution  indicating at the
  Fermi-level crossing.
However,  away from the (1,1) direction, as  shown by curves $\gamma_1$ and
 $\gamma_2$, the step in the
momentum distribution is smeared out by d-wave RVB order parameter,
 and the shadow band (the second peak beyond the Fermi surface)
 appears. Curve $\beta$ shows the momentum distribution along the
direction  $(\pi,0)$ to $(\pi,\pi)$, and there is no step detected,  indicating
 the absence of Fermi-level crossing in  that direction. The pseudo Fermi
surface
 is plotted in Fig. 3 for $\delta$=0.2, appearing as an unclosed small pocket.

We then calculate the physical electron spectral function in the
 $\alpha,\beta,\gamma_1,\gamma_2$
four directions.
The typical total spectral
weight of $\alpha$ and $\gamma_1$ cutes, including both coherent and incoherent
 parts,  is shown in Figs. 5-6.
 Fig. 5 shows the spectral weight along
the $\alpha$ cut ((1,1) direction). We  see the entire coherent part
 moving across the
Fermi level in this direction, and no signature of an energy gap
(depletion of spectral weight at the Fermi level) is noticed. In Fig. 6, the
pseudo-gap is visible at the Fermi level, and   the coherent part can still
move entirely across the Fermi level. Along the $\gamma_2$ direction
 the pseudo-gap is bigger than
 what we find in Fig. 6., which reflects the d-wave symmetry. On the
other hand, the spectral weight does not
drop to zero abruptly, decreasing continuously as the $k$ vector moves
along the $\gamma_2$ direction. Finally, for the $\beta$ cut
 we can hardly see any
 coherent part due to the  small value of the spinon distribution $n_s(k)$,
and the
spectral weight is
pushed down to the range much lower than the Fermi level, indicating at the
formation
of a maximal gap in this direction. Again, no abrupt Fermi-level crossing is
detected along this direction. Our result is fully consistent with the ARPES
 data obtained
by Loeser {\it et al.,}\cite{loe} which show  a nearly perfect Fermi-level
crossing
along the $\alpha$ cut and no Fermi level crossing along the
$\beta$ cut. They could also  find the pseudo-gap
from their experimental data, with vanishing value along the
$\alpha$ cut and a maximum along the $\beta$ cut. Similar results were
 also reported by H. Ding {\it et al.}\cite{ding}.

In Wen and Lee's approach\cite{wen}, the electron Fermi surface is
determined by the
pole of the renormalized  electron Green's function  due to
interaction with the gauge field.  An elliptical Fermi
surface with strongly suppressed spectral weight on the
 outer edge is obtained in the low doping regime.
This elliptic surface is not centered around $\Sigma$ point, which is on
the outer side and the wave function renormalization factor is exactly zero
at that point. Hence there is no "shadow band" in their approach
outside the boundary $(0, \pi)$ - $(\pi, 0)$. A spin gap  opens
outside the small elliptical Fermi surface, but the gap
is identically zero at the entire Fermi surface. Also, the quasiparticle
peak crosses the Fermi level in all directions spanned by the
Fermi surface.  The main features
of our results are similar to theirs, namely, small half-pocket-like
Fermi surface near $\Sigma$, and d-wave like gap in the
excitation spectrum. However, there are several important differences.
In our approach the spinon Fermi surface is adjusted by the
chemical potential, $i.e. $ the doping concentration, so the d-wave like
pseudo-gap is "locked" on the Fermi surface, and
the pseudo-gap  vanishes only at four
points at which the Fermi surface intersects  the line $k_x=\pm k_y$.
As soon as the points of the Fermi surface slightly deviate from the
$(1,\pm1)$ direction, a finite energy
gap appears. Also, the
quasiparticle peak of the spectral function does not cross
the Fermi level in any direction except for
 $(1,\pm1)$.
Another difference is we have small pockets centered around
$\Sigma$ and we do have shadow bands beyond the line
 $(0, \pi)$ - $(\pi, 0)$.  In principle, these different predictions can be
 checked by APRES experiments. It seems to us, however, that the
presently available resolution is still not enough to resolve this issue.

\vspace*{1cm}

\centerline{\bf  IV. CONCLUDING REMARKS}
\vspace{.5cm}

Using the slave-boson approach, within the mean field approximation,
 we have shown in this paper that
coexisting $\pi$-flux phase with d-wave RVB state is energetically
more favorable in the low doping range (below $\delta_c$ ),
compared with the pure d-wave RVB state. We have also shown
that this coexisting phase can provide a rather natural explanation
for the half-pocket-like small Fermi surface, a d-wave symmetry
pseudo-gap, and very different behavior for the spectral function in the
$\alpha$ and $\beta$ directions, in  underdoped HTSC. The main reason
for this outcome is we have been able to better treat the single-occupancy
constraint
by introducing the additional statistical gauge field.

The advantage of a mean field approach compared with more
sophisticated theories taking into account gauge
field fluctuations is that it gives a more transparent
understanding  of the experimental results. When studying the problem
of a single hole moving in the background of antiferromagnetic fluctuations,
an effective hole band is obtained by both numerical and analytic  methods.
The hole band  minimum is located at $(\pm\frac{\pi}{2},\pm\frac{\pi}{2})$.
Thus the Fermi
surface of holes in the very low doping regime should be small pockets
centered at
 $(\pm\frac{\pi}{2},\pm\frac{\pi}{2})$ due to  antiferromagnetic fluctuations.
This conclusion  is confirmed by the recent ARPES results. Hence the
antiferromagnetic fluctuations leading to a bipartial lattice
 must be considered in the low doping regime. In our
approach we introduce the $\pi$-flux phase which mimics the short range
antiferromagnetic correlations which  coexist with a weak d-wave RVB
phase. Intuitively,
the $\pi$-flux phase carries the physics of half-pocket-like Fermi surface
while the coexisting d-wave phase is responsible for the pseudo-gap opened
at the Fermi surface.

Zhang has concluded\cite{zhang}, however, that the $\pi$-flux state is
unstable
against the pure d-wave RVB state, excluding thus the possibility
of their coexistence, using a renormalized mean field theory within the
slave boson approach. What is the origin of the discrepancy between our
result and his conclusion?   We find that in
the mean field theory of the $t-J$ model the total energy is divided into
two parts.
One is the kinetic energy which is characterized by the normal average terms
$<f^+f>$ (for spinon) and $<h^+h>$ (for holons). The other is the exchange
energy which is characterized by the anomalous average term $<f^+f^+>$. The
$\pi$-flux state favors the kinetic energy, while the d-wave RVB state favors
the exchange energy. In Zhang's approach the dynamical and statistical
properties of holons which are shown to be nontrivial by our present study,
are not considered,   and the contribution of $<h^+h>$ is ignored.
This underestimates  the kinetic energy
gained by the delocalized motion of holons. Therefore the coexistence of $\pi$
flux and the d-wave RVB state is excluded in Zhang's calculation.
 In our present
study the dynamics of holons is fully considered, so  we find that in the
low doping
regime the $\pi$-flux state coexisting with d-wave phase is more stable
against both
the pure $\pi$-flux state and the pure d-wave RVB state.
The admixture of a small amount of the d-wave state
will provide holons with more energy gain at the Fermi surface  compared
to the pure $\pi$-flux phase. Due to the smallness of the Fermi surface
 in the coexisting state the d-wave
order parameter (of the order 0.1) is much smaller than the $\pi$-flux
order parameter
(of the order 1).  The
 energy gap  opens at the Fermi level due to the presence of the d-wave
RVB order parameter. Since there is no Bose condensation in our calculation,
this mixed state is responsible for the pseudo-gap in the normal state. With
the increasing of the doping concentration, the area enclosed by the spinon
 Fermi surface
is enlarged and the states near the Fermi surface take a higher percentage of
the total occupied states and play a more important
role compared with the low doping case. Therefore,  in the high doping range,
the d-wave RVB state gains free energy more effectively, and a crossover from
the coexistence state to a pure d-wave state takes place.
We believe our mean field treatment can serve as a good starting point for
further
improvement. It turns out that the mean field approximation of the
$U(1)\times SU(2)$
Chern-Simons bosonization approach for the 2D $t-J$ model\cite{marc} leads
to very similar results
as the present study.
The condensation of  holons will lead to superconductivity which was not
considered in this work. The fluctuations around the saddle point will create
an effective attractive interaction between spinons and holons, as recently
discussed  by Wen and Lee\cite{wen}. The role of these fluctuations in our
approach require further studies.

Very recently, Ding {\it et al.}\cite{ding1} reported that there seems to be
 a large Fermi
surface for HTSC over the entire range of doping, from underdoped to
overdoped cases. However, their resolution is still not satisfactory for the
substantially underdoped samples ($T_c= 15$K ). It is yet to be seen
 how strong is their evidence.
Also, the transport and optical data have shown rather strong evidence in favor
of small hole pockets in the highly underdoped samples\cite{ong}.



\newpage
\vspace{20mm}

\centerline{\large\bf   Figure Captions}
\vspace{1.0ex}
\parskip=2pt
\baselineskip=14pt

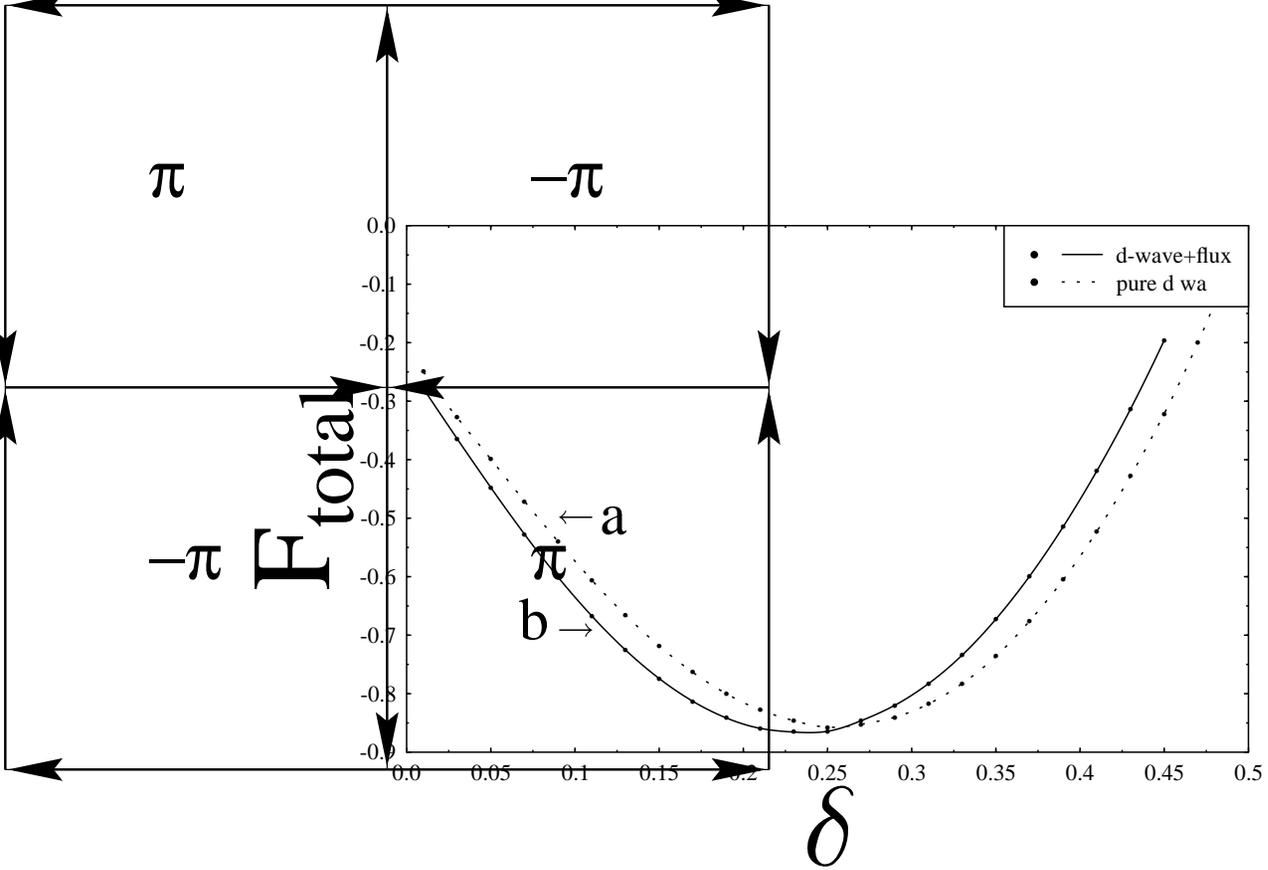
Figure~1.~Diagrammatic illustration of the $\pi$-flux  state. A
$\pi$($-\pi$) magnetic flux  flows alternatively through neighboring
plaquette of the lattice.

\vspace{2.0ex}

Figure~2.~The comparison of the total free energy F for the
coexisting $\pi$-flux + d-wave RVB
state and the pure d-wave RVB state.

\vspace{2.0ex}

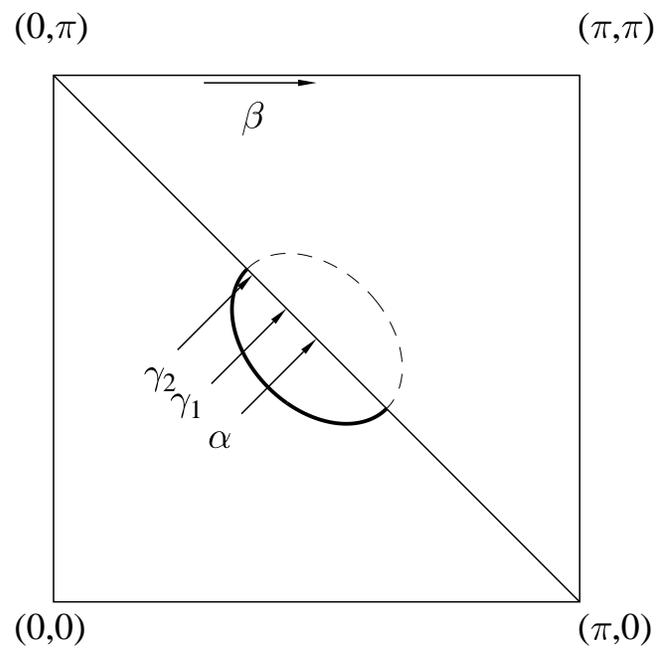
Figure~3.~The unclosed pocket-like spinon Fermi surface at
$\delta$=0.2. The four directions along which we calculate
the physical electron spectral function are  marked
in the figure as $\alpha$, $\beta$,$\gamma_1$ and $\gamma_2$.

\vspace{2.0ex}

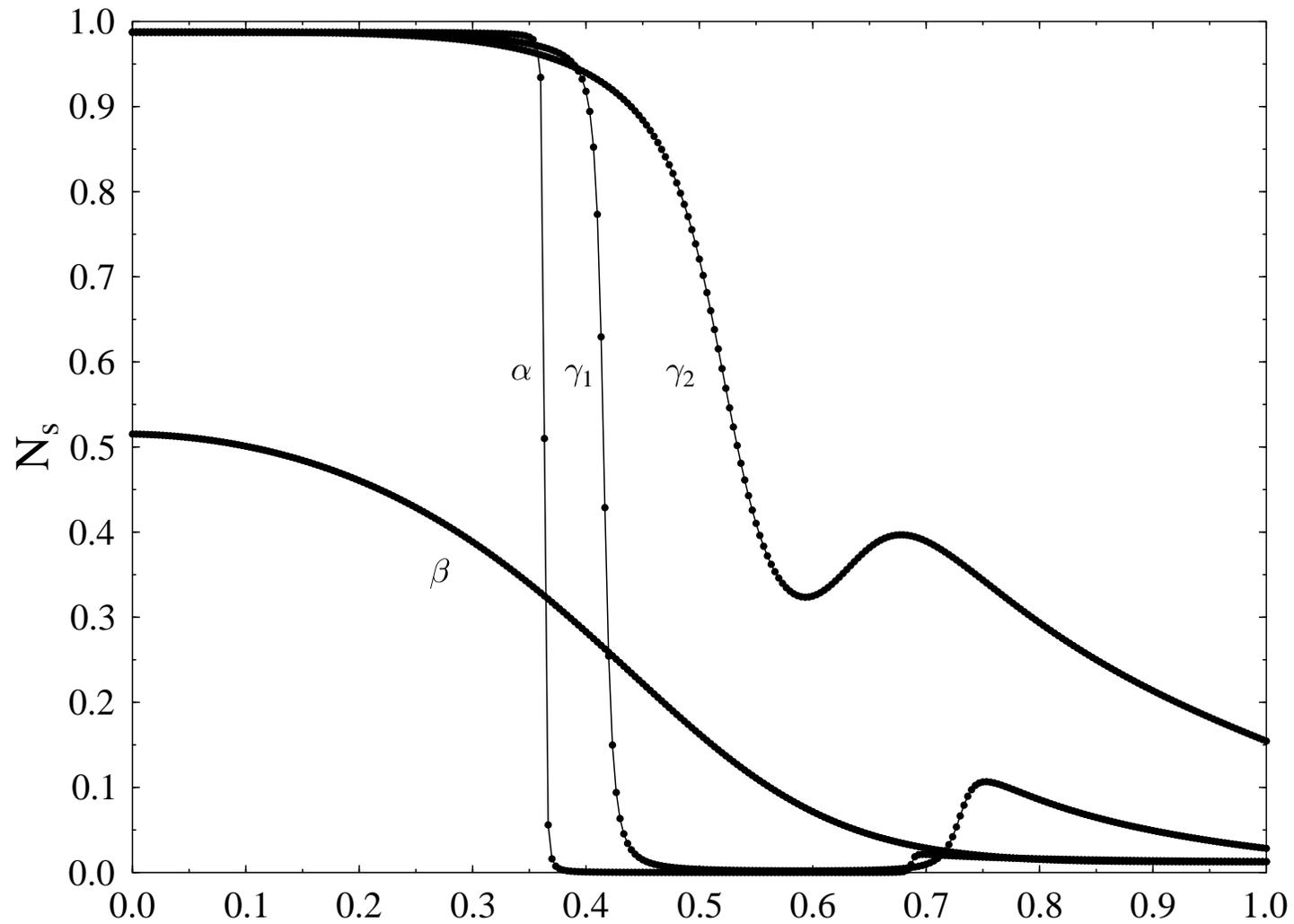
Figure~4.~The momentum distribution of the spinons along the four
directions defined  in Fig.3.

\vspace{2.0ex}

Figure~5. The physical electron spectral function along the $\alpha$
cut, from which we see a clear Fermi-level crossing and no pseudo-gap.
\vspace{2.0ex}

Figure~6. The physical electron spectral function along the $\gamma_1$
cut. A pseudo-gap opens, but the Fermi-level crossing is
still visible in this direction.

\vspace{2.0ex}

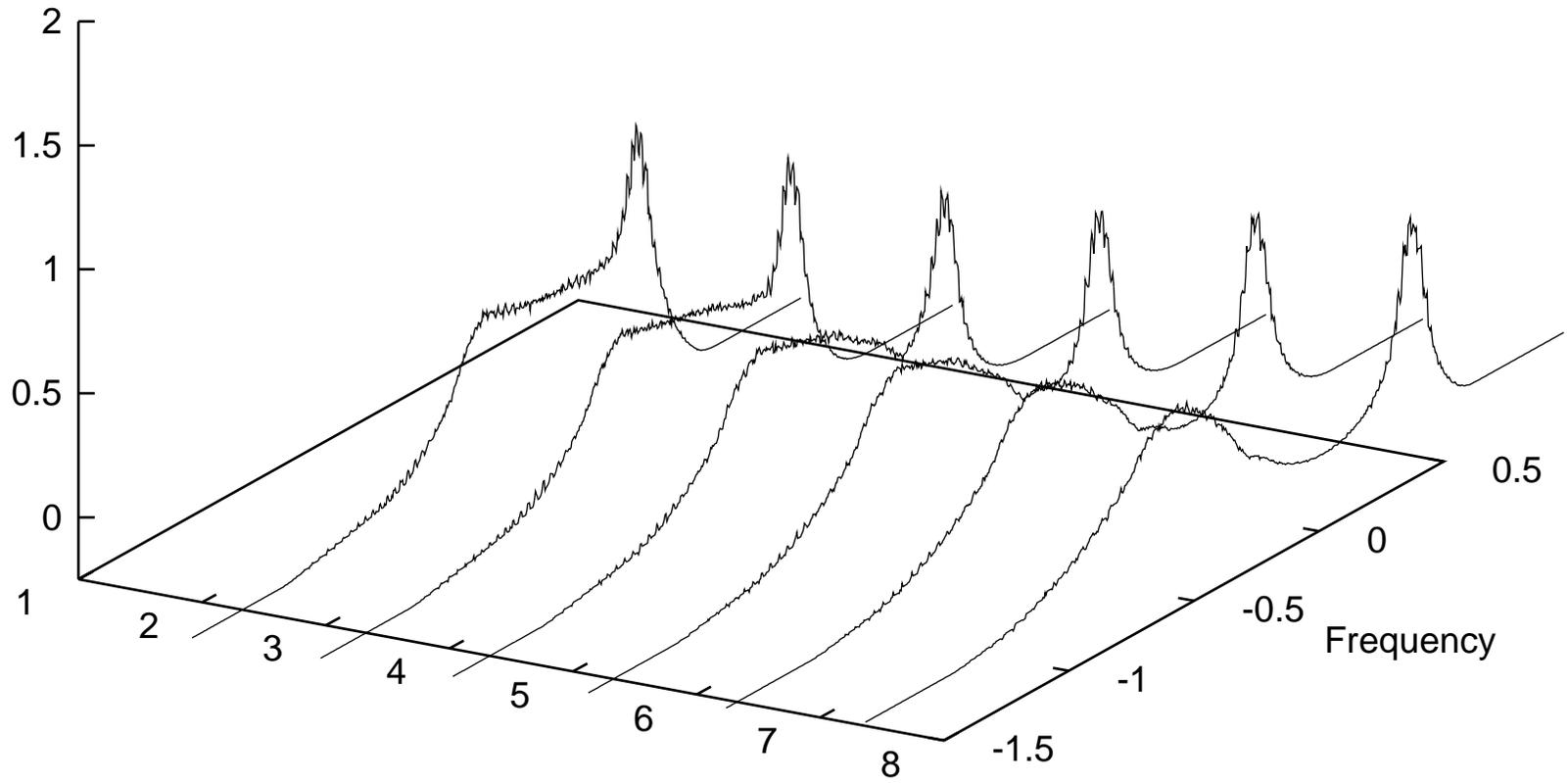
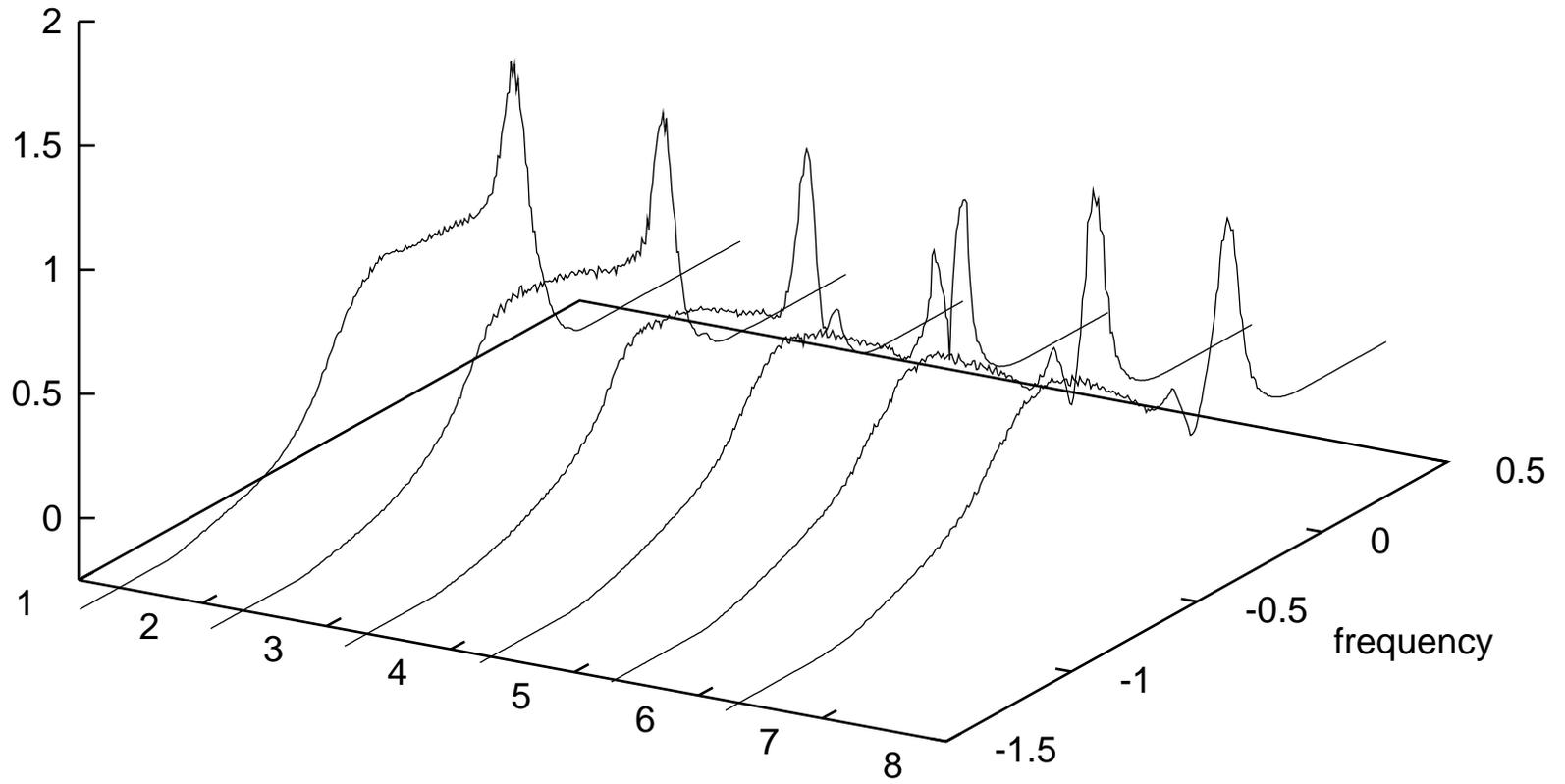
\end{document}